\documentclass[aps,prl,superscriptaddress,amsmath,amssymb,reprint,showkeys,footinbib,preprintnumbers]{revtex4-2}

\usepackage{amsmath}
\usepackage{graphicx}
\usepackage{bm}
\usepackage{xcolor}

\newcommand{\md}{\mathrm{d}}

\usepackage{amsfonts}   
\usepackage{amssymb}

\begin{document}
\preprint{}

\title{\textbf{Optimal navigation in a noisy environment} 
}

 \author{Abhijit Sinha} 
 \affiliation{Department of Condensed Matter Physics and Materials Science, Tata Institute of Fundamental Research, Homi Bhabha Road, Mumbai 400005, India.}
 \author{Sandeep Jangid}
\affiliation{Department of Theoretical Physics, Tata Institute of Fundamental Research, Homi Bhabha Road, Mumbai 400005, India.}
 \author{Tridib Sadhu}
\affiliation{Department of Theoretical Physics, Tata Institute of Fundamental Research, Homi Bhabha Road, Mumbai 400005, India.}
\author{Shankar Ghosh} 
 \affiliation{Department of Condensed Matter Physics and Materials Science, Tata Institute of Fundamental Research, Homi Bhabha Road, Mumbai 400005, India.}

\date{\today}

\begin{abstract}
Navigating toward a known target in a noisy environment is a fundamental problem shared across biological, physical, and engineered systems. Although optimal strategies are often framed in terms of continuous, fine‑grained feedback, we show that efficient navigation emerges from a far simpler principle: natural wandering punctuated by intermittent course corrections. Using a controlled robotic platform, active Brownian particle simulations, and scaling theory, we identify a universal trade‑off between noise‑induced deviation and the finite cost of reorientation, yielding an optimal course correction frequency governed by only a few system parameters. Despite their differing levels of complexity, our experiment and theory collapse onto common quantitative signatures, including first‑passage time distribution and non‑Gaussian angular dispersion. Our results establish intermittent course-correction as a minimal and robust alternative to continuous feedback, offering a unifying guiding principle for point‑to‑point navigation in complex environments.
\end{abstract}


\maketitle

Efficient navigation is a universal challenge across physical, biological, and engineered systems.
Pedestrians in corridors~\cite{bacik2025order,corbetta2017fluctuations}, ants on  trails~\cite{aguilar2018collective}, autonomously driven cars in traffic~\cite{schwarting2018planning}, animal migration~\cite{Hays2022TravelRoutesToRemote,Hays2020OpenOceanReorientation,Thorup2011JuvenileSongbirds,Horton2016Nocturnally,Merlin2012NavigationalStrategies,NaisbettJones2017MagneticMapEels}, and packets in congested networks~\cite{bertsekas2021data} all face the same fundamental problem: reaching a target while minimizing delays that incur costs in terms of time, energy, or throughput. A common strategy is to employ feedback mechanisms that limit wandering and reduce delays. While optimization strategies can be highly complex, multifaceted, and computationally expensive~\cite{Casert2024,DeBruyneMori2023Resetting,Putzke2023}, nature might rely on simpler principles to achieve near-optimal performance. This raises question: are there simple guiding rules for efficient navigation in complex environment?

Most navigation problems focus on target search, such as foraging~\cite{benichou2011intermittent,benichou2008optimizing}, searching lost objects~\cite{evans2011diffusion,evans2020stochastic}, or cells looking for pathogens~\cite{Meerson2015,Eisenbach2006}, where the target location is unknown, and movement is exploratory rather than goal directed navigation. In this letter, we address a fundamentally different scenario where fixed target location is known apriori, but the environmental factors drive an agent off course. This is a common scenario in many practical examples, such as targeted drug delivery~\cite{Fernandez-Rodriguez2020, Naahidi2013Biocompatibility}, commercial transport~\cite{11268231}, and robotics rescue missions~\cite{Albers2002Exploring}. In biology, cells follow chemical gradients towards fixed targets while facing a noisy environment~\cite{Swaney2010Eukaryotic,Levine2013PhysicsToday,Endres2008PNAS,Doitsidou2002Cell}. Animals migrate to a specific destination (breeding ground, food source) guided by inherited or learned map-and-compass mechanism~\cite{Hays2022TravelRoutesToRemote,Hays2020OpenOceanReorientation,Thorup2011JuvenileSongbirds,Horton2016Nocturnally,Merlin2012NavigationalStrategies,NaisbettJones2017MagneticMapEels}, while facing environmental uncertainty from atmospheric turbulence~\cite{Thorup2011JuvenileSongbirds}, variable magnetic cues~\cite{NaisbettJones2017MagneticMapEels}, or displacement events~\cite{Horton2016Nocturnally}. The problem of optimal navigation between two fixed points in a flow field is also important to broader domains of stochastic processes~\cite{Bartumeus2002,Stone1989} and relates to the Zermelo’s navigation problem~\cite{Shavakh2022Zermelo}.

Empirical evidence~\cite{AuthorLastname2025Title,Hays2020OpenOceanReorientation,NaisbettJones2017MagneticMapEels,McLaren2022Predicting,NaisbettJones2017MagneticMapEels} suggest a general rule of thumb: maintain a broad course with occasional coarse corrections rather than frequent fine adjustments. How often should a chaser correct its course to efficiently reach the target? Intuitively, frequent course corrections may seem to reduce travel time but incur costs due to reorientation delays arising from finite detection time or physical constrains such as retraction of motile protrusions before forming new ones~\cite{Romanczuk,ReichmanFried2004,EvansMajumdar2019}.

Our primary conclusion is that there exists an optimal course-correction frequency, determined by only a few system characteristics, that captures the essential navigation strategy without elaborate computational overheads. Accuracy of course correction is a secondary factor, unlike the emphasis of earlier studies~\cite{Romanczuk2015}.
This principle echoes intermittent search strategies and stochastic resetting~\cite{evans2011diffusion,fuchs2016stochastic,bechinger2016active,tal2020experimental,mendez2024first,biswas2025drift,santra2020run,mori2025optimal,besga2020optimal} in random target-search problems. 

Our conclusion emerges from a remarkable convergence of results across three approaches: controlled robotic navigation experiment, computer modeling based on
Active Brownian Particles (ABPs)~\cite{howse2007self,fily2012athermal,cates2015motility,bechinger2016active}, and scaling analysis. Our experimental platform retains key features of navigation problems — control costs, noise-induced deviations, and collision delays — while remaining simple enough for quantitative analysis, offering accessibility beyond scenarios such as human crowds, traffic, or animal collectives. The ABP-based computer model incorporates the inherent complexity of living matter due to additional scale from persistence length, in contrasts to most target-search studies done in Markovian, scale-invariant setting \cite{benichou2011intermittent,Reynolds2007}.
Despite their differing levels of complexity, the simple model and the scaling theory quantitatively captures several non-trivial experimental characteristics such as non-Gaussian angular dispersion, scaling of first-passage distributions, and the optimal course-correction frequency.
\begin{figure}
    \centering
    \includegraphics[width=1\linewidth]{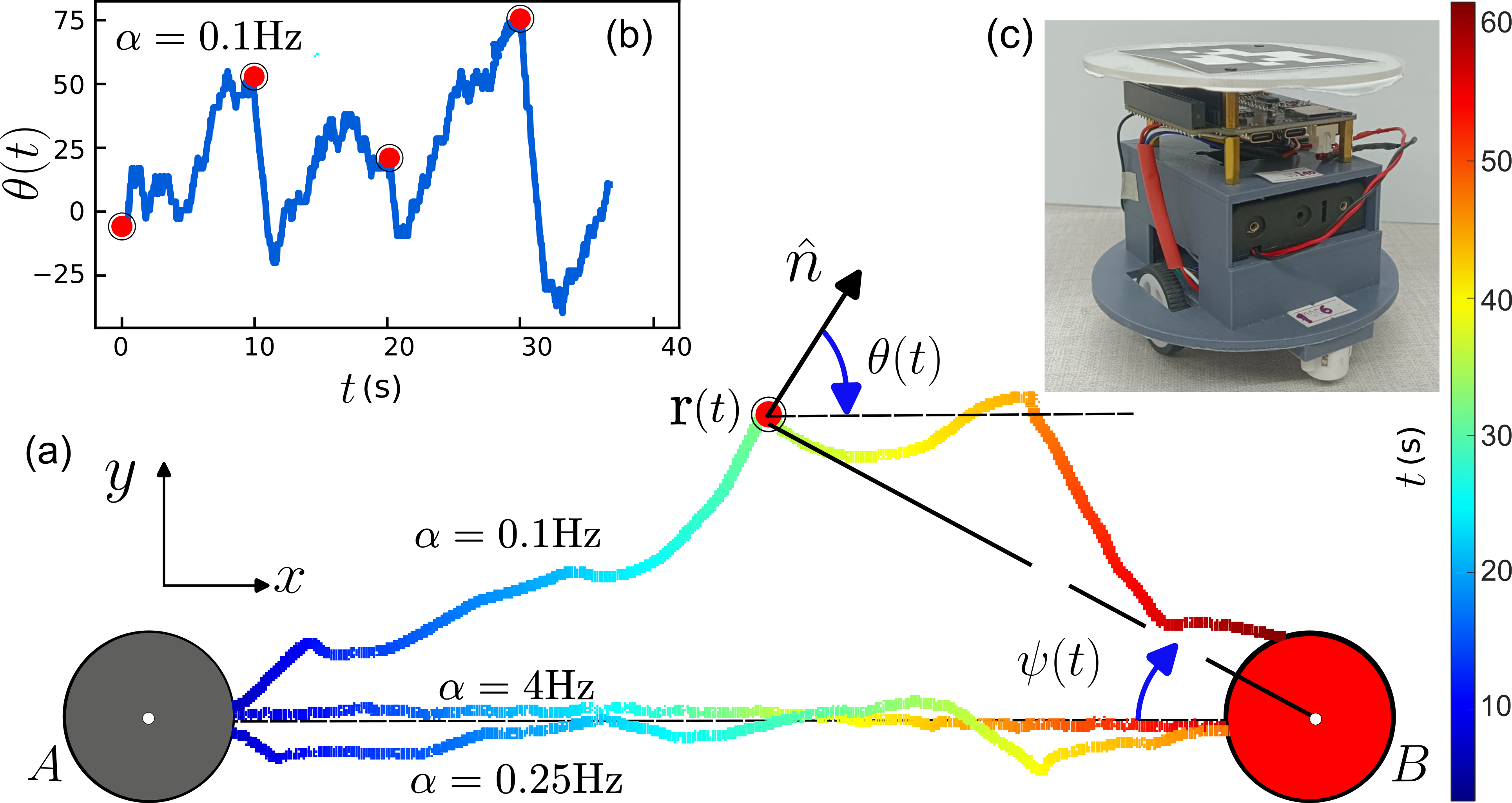}
  \caption{(a) Representative trajectories for different values of course-correction frequency $\alpha$ for a robot traveling from the 
$\epsilon$-neighborhood ($\epsilon=17~\text{cm}$) of region $A$ (grey circle) to that of region $B$ (red circle), 
separated by $AB\equiv L=150~\text{cm}$. At time $t$, the robot is at position $\mathbf{r}(t)$ 
with orientation $\theta(t)$ relative to the $x$-axis and heading vector 
$\hat{n}(t)=(\cos\theta(t),\,\sin\theta(t))$. The line of sight defines the angle $\psi(t)$, and at reset events at fixed intervals $1/\alpha$ the heading is realigned towards the target, 
i.e. $\theta(t)\to -\psi(t)$. The robot is considered to have reached the goal 
once it  enters the $\epsilon$-neighborhood of $B$. The color gradient (blue to red) of the trajectories indicate the passage of time, including time delays from angular reset.  (b)
A sample evolution of the orientation $\theta$ (in degrees) with time $t$ (in seconds) for $\alpha=0.1$Hz. The red points mark the initiation of the resetting events. (c) Picture of differential-drive 
robot used in the experiment.
}
    \label{fig:setup}
\end{figure}

Our experimental navigation setup (see Fig.~\ref{fig:setup}) involves a single robot moving between $\epsilon$ neighborhood of two fixed points, $A$ and $B$, separated by a distance $L$.
We employ a differential-drive robot~\cite{siegwart2011introduction} 
mounted on a custom 3D-printed chassis (Fig.~\ref{fig:setup}c) actuated by two independently driven DC motors with encoders and powered by an ESP32 micro-controller. Wheel motion is controlled through pulse-width modulation (PWM), producing an average linear velocity $v_0$. 
Real-time communication with a host PC is established via Wi-Fi, enabling remote navigation and continuous data acquisition. Motion and power consumption are tracked using wheel encoder signals in combination with a current sensor~\cite{INA219}. 

However, as is often the case with off-the-shelf components, the two motors are not perfectly matched. One motor consistently performs better than the other, leading to a slight but persistent drift in the robot’s path. Thus, rather than trying to perfectly calibrate the motors, which can be tedious and unreliable, we take a different approach. We introduce a controlled amount of randomness into the system. The robot alternates between activating the left and right motors in short bursts, creating a kind of “wiggling” motion. This deliberate noise helps cancel out the bias caused by motor mismatch. Over time, this randomized movement mimics a random walk in orientation, allowing the robot to maintain a rough straight trajectory with a fixed speed $v$ up to a characteristic persistent time. An overhead camera (Pixelink-PL-D734CU-T) tracks the robot in real time, using AprilTags~\cite{wang2016apriltag} to extract precise position and orientation data. 

In our set up the robot aims to travel from A to B, and vise verse. Such a navigation problem inevitably involves uncertainties, which may arise from noisy sensors and environmental variability. A traditional approach~\cite{thrun2002probabilistic} treats the uncertainties as stochastic processes and makes continuous adjustments using updates from the sensors requiring significant computational resources.
We introduce different approach, where using a minimal feedback system  the robot's heading is periodically reset at fixed time intervals $1/\alpha$ toward its target.

\begin{figure}
\centering
\includegraphics[width=1.0\linewidth]{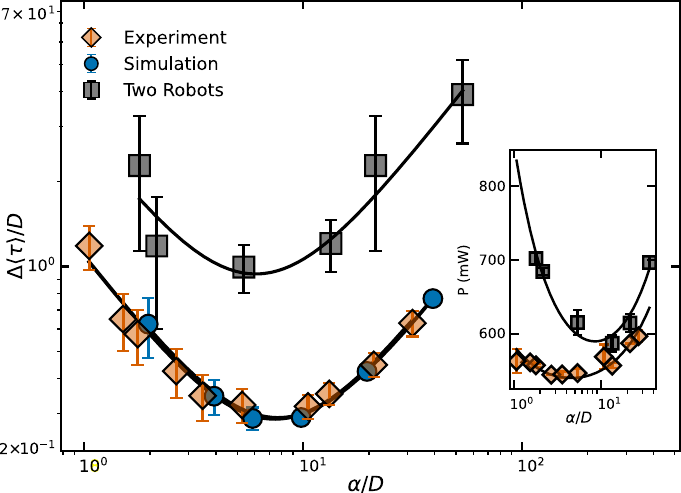}
\caption{Normalized mean target-hitting time $\Delta\langle \tau \rangle/D$ versus $\alpha/D$ for the experiment and the ABP-simulation. The experimental results are for the single-robot and the double-robot problems. The solid lines show non-monotonic dependence of the form in \eqref{eq:mfpt_rescaled}. The inset shows the mean power consumption with the solid lines representing an empirical fit
$P_0 + P_1/\alpha + P_2 \alpha$, with 
($P_0=526,\; P_1=43,\; P_2=2.6$) for the single-robot case and ($P_0=509,\; P_1=344,\; P_2=4.7$) the two-robot problem.}
\label{fig:MFT}
\end{figure}

Fig.~\ref{fig:setup}a shows a typical trajectory of a robot moving from $A$ to $B$.  At periodic intervals \cite{besga2020optimal,Altshuler2024} set by the reset frequency $\alpha$, the robot is halted and reoriented until its heading lies within $\pm 5^{\circ}$ of the target. Kinks, indicated by red dots in the evolution of angle in Fig.~\ref{fig:setup}b shows the re-orientation events. Frequent re-orientations suppress large deviations, confining trajectories to a narrow region, while rare resets allow broader wandering.  

There is a trade-off between keeping the robot on course and the associated cost of time delay in reorientation. Frequent course correction improves accuracy but at the cost of time delay, whereas less frequent resets allow movement noise to accumulate, increasing the travel time. An intermediate reset frequency leads to lesser travel time as illustrated in the passage of time in the three sample trajectories in Fig.~\ref{fig:setup}a: the $\alpha=0.25Hz$ trajectory is fastest among the three samples.  

We report a non-monotonic relationship between the course-correction frequency $\alpha$ and the mean travel time $\langle \tau \rangle$ to reach the target. In Fig.~\ref{fig:MFT} we plotted the relative excess time $\Delta \langle\tau \rangle =\frac{\langle \tau \rangle}{\tau_{\rm bllst}}-1$ over the head-on flight time $ \tau_{\rm bllst}=\frac{L-2\epsilon}{v}$ where $v$ is the speed of the robot (see End Matter).
Our data suggest a simple dependence 
\begin{equation}
\frac{\Delta \langle \tau \rangle}{D}
= \frac{1}{\alpha}
+ \tau_{0} \frac{\alpha }{D}
    \label{eq:mfpt_rescaled}
\end{equation}
on $\alpha$, the average time-lag $\tau_0$ for reorientation, and the angular diffusivity $D$ of the free evolution.
In \eqref{eq:mfpt_rescaled}, the $1/\alpha$ term reflects the suppression of angular wandering by frequent resets, while the linear term captures the cumulative overhead for reorientation. Their competition gives a well-defined optimal course-correction frequency
$\alpha_{\mathrm{opt}} = \sqrt{D/\tau_{0}}$. At low $\alpha$, course-corrections are beneficial because they correct angular drift, whereas at high $\alpha$, the growing time cost of frequent reorientation outweighs this advantage, making the process inefficient. The divergence in \eqref{eq:mfpt_rescaled} for vanishing $\alpha$ contrasts $1/\sqrt{\alpha}$ dependence in positional resetting \cite{Baouche2025,evans2020stochastic}.

In practice, the time-lag $\tau_0$ of reorientation depends on the amount of angular change. Larger reorientation takes longer time. Considering the angular standard deviation $\sqrt{2D/\alpha}$ at reset events at $1/\alpha$ intervals, we get
\begin{equation}
\tau_{0}
= \tau_{1} + \frac{\tau_{2}}{\sqrt{\alpha/D}
}
\end{equation}
where $\tau_1=0.14$ and $\tau_2=0.08$ are independently measured, along with angular diffusivity $D=0.09$ (see End Matter). The non-zero $\tau_1$ is the cumulative overhead from delays due to communication, information processing, etc.

When multiple robots operate in the same arena, an additional contribution arises: the time lost to resolving collisions. These encounters prolong the first-passage time. High resetting frequencies promote head-on collisions, which take longer to resolve, while lower resetting frequencies favor grazing encounters that are quicker to resolve but occur along longer paths.  
Despite the added complexity introduced by these interactions, the data in Fig.~\ref{fig:MFT} for the two robots case  remains well-described by a modified form of the single-robot model:$\frac{\Delta \langle \tau \rangle}{D} = \left(\frac{\beta}{\alpha}\right) + \tau_0' \left(\frac{\alpha}{D}\right)$, where  the fitted parameters $\beta \approx 2.4$ and $\tau_0' \approx 0.8$ account for the time penalties associated with collision avoidance.

The non-monotonic dependence is also prevalent in the average power consumed per robot.  In our experiment, we measure time series of the current $I$ and voltage $V$, and calculate the mean electrical power from the battery as the time average of their product, $P=\langle IV\rangle$. The inset of Fig.~\ref{fig:MFT} shows $P(\alpha)$ for one and two robots problems which fits well to $P_0 + \frac{P_1}{\alpha} + P_2 \alpha$.  While purely empirical, its structure is consistent with a baseline locomotion power consumption $P_0$, a term decreasing with $\alpha$ associated with energy cost from wandering, and a linear term $P_2\alpha$ coming from reset overheads. The latter comes from surge in power consumption for generating slips from counter-rotating wheels to reset heading angle, sensing, and communication. The linear growth is not a trivial consequence of the variation in passage time, since $P$ already represents energy per unit time and does not scale with the duration; 
rather, it reflects how power is spent as a function of the reset rate.

\begin{figure}[t]
    \centering
    \includegraphics[width=0.95\linewidth]{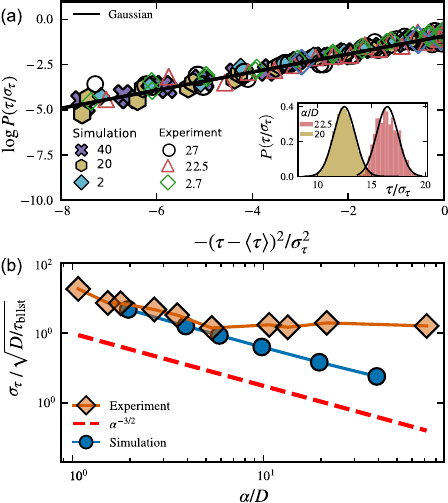}
      \caption{(a) Logarithm of  normalized histogram for rescaled target-hitting time $P(\tau/\sigma_\tau)$ versus $-(\tau-\langle\tau\rangle)^2/\sigma_\tau^2$ for different $\alpha/D$, where $\sigma_\tau$ is the standard deviation of $\tau$. Data collapse to a straight line confirms a Gaussian distribution. Inset: the corresponding histogram in linear scale, from experiment ($\alpha/D=22.5$) and simulation ($\alpha/D=20.0$), with a gaussian fit indicated by the solid curves.  
(b) Normalized standard deviation as a function of $\alpha/D$ on a log-log plot. The dashed line indicates a power-law $(\alpha/D)^{-3/2}$.
 }
    \label{fig:time_distribution}
\end{figure}

The dependence \eqref{eq:mfpt_rescaled} is robust and quantitatively emulated even in a simple toy model: ABP with directional resetting. The position $\mathbf{r}$ for ABP ~\cite{howse2007self,fily2012athermal,cates2015motility,bechinger2016active} in two-dimensions evolves at a constant speed $v$, 
$\dot{\mathbf{r}} = v(\cos\theta,\, \sin\theta)$
with heading angle $\theta$ following unbiased diffusion with angular diffusivity $D$. 
In addition, the angle $\theta$ is periodically reset at uniform intervals $1/\alpha $ towards the fixed target, by sharply changing the angle $\theta \to -\psi$ where $\psi$ is defined in Fig.~\ref{fig:setup}a. Similar directional resetting has been studied \cite{Shee2024,Kumar2020,Baouche2025,Basu2024,Romanczuk2015,Paramanick2024} in related context.

This simple dynamics surprisingly captures more intricate dynamics of the robots, beyond the qualitative physics. The target-catching time from direct numerical simulation of the model, with corresponding parameter values $D$ and $v$ from the experiment, is shown in Fig.~\ref{fig:MFT}, which compares extremely well with the corresponding experimental result, reflecting universality in this navigation strategy.

The dependence \eqref{eq:mfpt_rescaled} is  derived in the End Matter for this toy model. The non-trivial $1/\alpha$ dependence in \eqref{eq:mfpt_rescaled} is naively understood from the effective dynamics in the fast-reset regime. Between resets, angular diffusion produces small transverse deviations that lengthen the path. Since the inter-reset interval is $1/\alpha$, the accumulated excess traveled distance between resets decreases as $1/\alpha$, giving the first contribution in \eqref{eq:mfpt_rescaled}.

The quantitative overlap between experiment and theoretical model, also extends for the distribution of the target catching time. Figure~\ref{fig:time_distribution}a shows the Gaussian probability distribution of the rescaled first-passage time, $P(\tau/\sigma_\tau)$, for different $\alpha/D$, where $\sigma_\tau$ is the standard deviation of $\tau$. There is an excellent collapse of data from both experiment and simulation onto a Normal distribution. The distribution simply emerges (see End Matter) from central limit theorem applied to successive reset intervals and contrasts slower exponential tail in Brownian resetting \cite{evans2020stochastic}.

The variability of $\tau$ is measured by the normalized standard deviation, 
$\sigma_\tau\sqrt{D / \tau_{\rm  bllst}}$, which decays as $(\alpha/D)^{-3/2}$ (Fig.~\ref{fig:time_distribution}c), in good agreement between experiment and the model simulation. The deviations of the experimental data from this trend at large $\alpha$ is due to imperfect alignment in resetting events. In practice, the robot heading is reset within $\pm 5^{\circ}$ of the line of sight towards the target, imposing a lower bound on the standard deviation.

The scaling of $\sigma_{\tau}$ with $\alpha$ can be rationalized by considering the fast-resetting regime, where the particle makes many short, nearly straight runs toward the target. Each run is almost ballistic, but angular noise introduces small sideways deviations that slightly lengthen the path. The size of these fluctuations decreases as the reset frequency increases, while the number of reset steps grows proportionally with $\alpha$. The interplay of these two effects---diminishing fluctuations per step but more frequent steps overall---leads to fluctuations in first-passage time that decay as $\alpha^{-3/2}$ (see End Matter).

\begin{figure}
       \centering
       \includegraphics[width=1.0\linewidth]{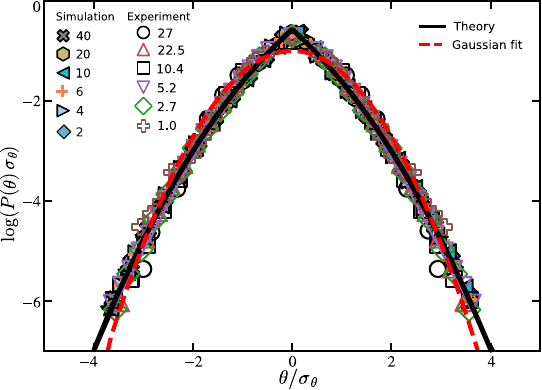}
         \caption{Distribution of angle $\theta$, rescaled by its standard deviation $\sigma_\theta$, for different values of rescaled reset frequencies $\alpha/D$ indicated in the legends. The solid line is the non-Gaussian distribution in \eqref{eq:theta_distribution 1}, while the dashed line shows a Gaussian fit. The scaling collapse of all data, experimental and numerical, demonstrates a universal angular statistics that evades central limit regime.}
     \label{fig:angle_histogram_expt}
    \end{figure}

The directional resetting introduces a non-trivial angular dispersion. The distribution of the angle $\theta$, measured stroboscopically during the travel time, is shown in Fig.~\ref{fig:angle_histogram_expt}. For quantitative comparison of experimental and ABP-model data, the angles are rescaled by their respective values of standard deviation $\sigma_\theta$. The near-perfect scaling collapse  shows $P(\theta)= \frac{1}{\sigma_\theta}f\left(\frac{\theta}{\sigma_\theta} \right)$ with a smooth function
\begin{equation}
    f(x)=\frac{e^{-\frac{x^2}{4}}}{\sqrt{\pi}}-\frac{\vert x \vert}{2} \textrm{erfc}\left(\frac{\vert x \vert}{2}\right)\label{eq:theta_distribution 1}
\end{equation}
derived from the ABP-dynamics (see End Matter). Small fluctuations have a sub-Gaussian Laplace distribution ($e^{{-\vert x \vert }}$) with a singularity at the origin, thus not describable by the central limit theorem. This suppression of small fluctuations originate from the confining influence of angular reset. Large fluctuations follow a conventional Gaussian decay. The distribution \eqref{eq:theta_distribution 1} was also reported for resetting gas \cite{BiroliMajumdarSchehr2025}, while the non-gaussian fluctuations ($e^{{-\vert x \vert }}$) was known for the stationary distribution of resetting Brownian \cite{Majumdar2015Resetting,yerrababu2024}.  The gaussian asymptotics for large fluctuations is a consequence of our periodic resetting protocol \cite{Kumar2020,bechinger2016active,cates2015motility} compared to the usual stochastic protocol \cite{Majumdar2015Resetting,yerrababu2024}.

A few remarks are in order.
Despite inherent complexity of the robots, the resetting  scenario is quantitatively described by our simple toy model and scaling analysis (see End Matter) indicating towards an universal navigation strategy.  

For the robot navigation, the re-orientations were never perfect and this error reveals in the distribution of target-hitting time in Fig.~\ref{fig:time_distribution}. However, its effect in the mean target hitting time is negligible, as seen in Fig.~\ref{fig:MFT} from the quantitative description of ABP where re-orientations are perfect. This confirms that, for optimal navigation, course-correction frequency is the leading control factor.

The optimal average target-catching time in Fig.~\ref{fig:MFT} corresponds to a reset frequency $\alpha>D$. This suggest a rule of thumb to correct course frequently enough compared to persistence time ($1/D$). 

In conclusion, we have introduced a directional resetting framework for  navigation to a fixed target, highlighting the similar essential trade-off previously emphasized in intermittent search strategies: resets confine noise-induced wandering but incur overheads, giving rise to an optimal reset frequency. Compared to traditional resetting paradigm~\cite{evans2011diffusion,tal2020experimental,mendez2024first,biswas2025drift,santra2020run,mori2025optimal,besga2020optimal} in random target search, our contribution shifts the perspective to target navigation where both sensing and control are costly. In this setting, periodic interventions emerge as an efficient strategy to stabilize dynamics without continuous feedback.  
The quantitative description in terms of simple model suggest a universal characteristic of this strategy. 
Our work provides a guiding principle that would be relevant for controlling living matter \cite{Casert2024,DeBruyneMori2023Resetting,Putzke2023,Liebchen2019Optimal}, with applications in design of ``smart active matter'' \cite{Levine2023,Casert2024}, bio-inspired technologies, such as micro-robotic drug delivery \cite{Naahidi2013Biocompatibility}, cargo transport \cite{Palagi2018Bioinspired}, and robotics for conducting life saving rescue missions \cite{Albers2002Exploring}.

\begin{acknowledgments}
\textit{Acknowledgments}: We thank Anit Sane for technical support with the experimental setup. A.S. and S.G. acknowledge the financial support of the
Department of Atomic Energy and Science and Engineering
Research Board, Government of India, under Projects No. 12-
R\&D-TFR-5.10-0100 and No. CRG/2020/000507. TS acknowledges the financial support of the Department of Atomic Energy, Government of India, under Project Identification No.~RTI 4002. TS
thanks the support from the International
Research Project (IRP) titled `Classical and quantum dynamics in out of equilibrium systems' by CNRS, France. 
\end{acknowledgments}

\section{End Matter}

\section*{Experimental Setup and Methods}

\paragraph{Experimental Setup and Robot Architecture:} Experiments were conducted in a $50 \times 100$ inch wooden arena with the robot's pose tracked by an overhead Pixielink camera using AprilTags mounted on the chassis. The camera was geometrically calibrated using a checkerboard and captured high-resolution images ($800 \times 1400$ pixels). To maintain real-time performance, full-frame AprilTag detection was used only for initial localization; subsequent tracking relied on a dynamically updated region of interest (ROI) centered on the robot's estimated position. The robot itself is a differential-drive mobile platform built on a compact 3D-printed chassis. Locomotion is powered by two N20-type geared DC motors equipped with Hall-effect quadrature encoders, allowing for closed-loop measurement of wheel rotation. The wheel radius is $R=17$ mm and the axle length (distance between centers of two wheels) is $L_{\mathrm{axel}}=75$ mm. Control is managed by a Waveshare ESP32-based controller board powered by three 3.7 V Li-ion cells (11.1 V nominal). Motor velocities are regulated via PWM.The control software utilizes a modular, multi-threaded architecture to separate time-critical tasks like vision processing and state estimation from less critical functions like data logging and communication. The robot firmware, structured as parallel FreeRTOS tasks, handles motor actuation, odometry, power sensing, and WiFi communication, enabling deterministic execution of motion primitives (e.g., forward motion, turning, and the randomized ``wiggle" routine). Full details on the robot's construction are provided in the Supplementary Information (SI).

\paragraph{Stochastic Wiggle Mode, Angular Diffusion, and Resetting Dynamics:} The robot achieves forward movement using a ``wiggle mode" implemented as a minimal stochastic control policy: at fixed intervals $\epsilon$, the robot randomly actuates either the left or right wheel with equal probability. This mechanism generates small, unbiased angular increments ($\delta\gamma_i$) in the heading, leading to diffusive orientation dynamics. This heading evolution is characterized as a random walk, quantified by the effective angular diffusion constant ($D$):
$$D = \frac{\langle \gamma^2 \rangle}{2t} = 2\epsilon\left(\frac{R\omega_0}{L_{\mathrm{axel}}}\right)^2$$
Using experimentally measured parameters, the diffusion constant is calculated as $D = 0.09$. The experimentally measured mean characteristic resetting time ($\langle \tau_0 \rangle$)  fits the functional form:$$\langle \tau_0 \rangle = \tau_1 + \frac{\tau_2}{\sqrt{\alpha/D}}$$Here, $\tau_1$ is the constant reorientation overhead, and $\tau_2$ is a characteristic timescale set by the diffusive angular wandering. The effects of stochastic resetting are evident in the robot's transverse motion: the Mean Squared Displacement (MSD) saturates at a plateau level that decreases as the resetting rate ($\alpha$) increases, and the variability in sideways distance (standard deviation, $\sigma_{y}$) scales inversely with the resetting rate ($\sigma_{y} \propto 1/\alpha$), confirming that faster resetting significantly constrains transverse spatial spread. Full details regarding the diffusion constant, the resetting-time scaling, and the transverse motion characteristics are provided in the Supplementary Information (SI).

\paragraph{Power Consumption Characterization:} The robot's power consumption was characterized to quantify the energetic cost of four operational modes (Idle, Wireless Communication, Wiggling, and Turning in Place) and to calibrate the onboard INA219 current sensor. Voltage and current were measured at $10\text{ Hz}$ using calibrated Keithley multimeters connected to an 18650 Li-ion battery. Analysis of the power probability density profiles  revealed distinct mean consumption values: Idle (0.405 W), Communication (0.869 W), Wiggling (1.054 W), and Reorientation (1.471 W). This profile confirms Reorientation is the most energetically costly mode and provides the necessary data for sensor calibration. Detailed power histograms are available in the Supplementary Information (SI).

\paragraph{Locomotion Dynamics and Sources of Stochastic Uncertainty:} The robot's pose, $\mathbf{x}_t = [\,x_t,\; y_t,\; \theta_t\,]$, evolves between resets via odometry, where the control inputs $\mathbf{u}_t = [\,v_t,\;\omega_t\,]$ are derived from the left ($\omega_L$) and right ($\omega_R$) wheel angular velocities. Specifically, $v_t = \frac{r}{2}(\omega_R+\omega_L)$ and $\omega_t = \frac{r}{L_{\mathrm{axel}}}(\omega_R-\omega_L)$. The locomotion variability is defined by two error types: Process Noise ($Q_t$), which accumulates continuously during odometry ($\mathbf{x}_{t+1} = f(\mathbf{x}_t,\mathbf{u}_t) + Q_t$), leading to wandering behavior; it arises from the deliberate ``wiggle" trajectory and actuation uncertainties, which are highlighted by the spread in the $\omega_L$ and $\omega_R$ speed distributions . The other is Measurement Noise ($R_t$), which enters discretely at AprilTag-based reset events; this noise stems not from detection error but from the pre-defined tolerance for heading realignment, requiring the orientation to be within $\pm 5^\circ$ of the target to complete the reset, making the $\pm 5^\circ$ residual misalignment the dominant measurement uncertainty. The histograms of $\omega_L$, $\omega_R$, $v_t$, and $\omega_t$ are provided in the Supplementary Information (SI).

\paragraph{Head-on Collisions Between Robots:} 

When multiple robots operate simultaneously under the resetting protocol,
trajectory intersections lead to collision events that interrupt forward motion
and introduce an additional delay in transport. We analyzed binary collisions
between robots executing identical reset dynamics to quantify this effect.

Collisions were identified when the inter-robot distance fell below a threshold
comparable to the robot diameter. Each event was characterized by an entry angle
$\beta_{\mathrm{in}}$, describing the approach geometry, and an exit angle
$\beta_{\mathrm{out}}$, describing the separation geometry, both defined with
respect to the line of centers. Small $\beta_{\mathrm{in}}$ corresponds to
head-on encounters, while large $\beta_{\mathrm{in}}$ indicates grazing
collisions.

Collision statistics depend strongly on the resetting rate~$\alpha$. Increasing
$\alpha$ enhances the probability of head-on encounters, while the exit
geometry remains largely insensitive to~$\alpha$, indicating that separation
dynamics are governed primarily by local avoidance interactions. Head-on
collisions are both more frequent and longer-lived at higher~$\alpha$, whereas
grazing encounters resolve rapidly. Consequently, increasing~$\alpha$
effectively increases the delay associated with each reorientation event. Within
the mean first-passage framework, this appears as an upward renormalization of
the reorientation time~$\tau_0$, shifting both the position and depth of the
minimum in $\langle \tau \rangle(\alpha)$. Detailed definitions, angle
distributions, and collision residence-time statistics are provided in the
Supplementary Information.

\section*{Active Brownian particle with directional resets}

We model the robot as an active Brownian particle (ABP) moving in two dimensions with constant speed $v$, with orientation $\theta(t)$ undergoing rotational diffusion according to
\begin{equation}
    \dot{x}=v\cos\theta,\quad 
    \dot{y}=v\sin\theta,\quad
    \dot{\theta}=\sqrt{2D}\,\eta(t),
    \label{eq:ABP_dynamics}
\end{equation}
where $\eta(t)$ is Gaussian white noise and $D^{-1}$ is the persistence time. For $t\ll D^{-1}$ the motion is ballistic, while for $t\gg D^{-1}$ the heading fully decorrelates and the movement follows normal diffusion. For reaching a fixed target, the particle starts facing the target, and over time, as its wanders off course, the heading is reset periodically at fixed intervals $\Delta t=\alpha^{-1}$ towards the target. 

For analyzing the target arrival time, we focus on the limit $\alpha\gg D$, where reset interval $\Delta t$ is much smaller than persistence time so that between successive resets the particle moves almost in a straight line. Moreover, the angle $\theta$ remains small in this limit. 

Over such an interval between two successive resets, the displacement $\Delta x\simeq v\Delta t$ and $\Delta y\simeq v\Delta t\,\Delta\theta$ with $\Delta\theta\simeq \sqrt{2D}\,\Delta w$, where $\Delta w$ is a random variable with $\langle\Delta w\rangle=0$ and $\langle\Delta w^2\rangle=\Delta t$. The total path length traveled between successive resets,
\[
R\simeq \sqrt{\Delta x^2+\Delta y^2}\simeq \Delta x +\frac{\Delta y^2}{2\Delta x}.
\]
If there are total $n$ reset events before reaching the target, the total traveled distance
\[
R_{\rm tot}\simeq \ell + D v\Delta t\sum_{i=1}^n\Delta w_i^2 ,
\]
where $\ell$ is the head-on distance to the target. 
Considering that the particle travels almost in a straight line between resets, the total time of flight $\tau_{\rm flt}=R_{\rm tot}/v$. Each reset requires a finite reorientation time $\tau_0$, and combining these we get the time to reach the target
\begin{align}
    \tau\simeq\frac{\ell}{v}
    +D\Delta t\sum_{i=1}^n\Delta w_i^2
    +n\tau_0.
    \label{eq:FPT}
\end{align}
Its average
\[
\langle\tau\rangle\simeq\frac{\ell}{v}
+\frac{D}{\alpha^2}\langle n\rangle 
+\langle n \rangle\tau_0.
\]
where we used $\langle \Delta w^2\rangle=\Delta t$ and $\Delta t=1/\alpha$. To leading order, $\langle n \rangle \simeq  \alpha \ell/v$ and this gives the excess time
\[
\Delta\langle\tau\rangle=\langle\tau\rangle-\tau_{\rm bllst}
\simeq \left(\frac{D}{\alpha}+\tau_0\alpha\right)\tau_{\rm bllst},
\]
where $\tau_{\rm bllst}=\ell/v$, leading to \eqref{eq:mfpt_rescaled}.

We also obtain the variance from \eqref{eq:FPT} using central limit theorem,
\[
\langle\tau^2\rangle_c \simeq D^2 \Delta t^2 n \langle (\Delta w^2)^2 \rangle_c=2D^2 \Delta t^4 n= 2D^2 \frac{\tau_{\rm bllst}}{\alpha^3}.
\]
The central limit theorem also suggest that the distribution of $\tau$ in \eqref{eq:FPT} is Gaussian, 
\[
P(\tau)=\frac{1}{\sqrt{2\pi\sigma_\tau^2}}
\exp\!\left[-\frac{(\tau-\langle\tau\rangle)^2}{2\sigma_\tau^2}\right],
\]
with $\sigma_\tau^2=\langle\tau^2\rangle_c$. Both the distribution and the scaling of the variance are confirmed in Fig.~\ref{fig:time_distribution}.

\paragraph{Angular dispersion:}  
For deriving the angular statistics under directional reset, we consider first a simpler problem of a one-dimensional Brownian particle $X_t$ subject to periodic reset to its initial position $X=0$ at intervals $\Delta t =1/\alpha$. Between two successive resets, the particle follows free Brownian motion. The time-dependent probability distribution 
\begin{equation}
    P_t(x) = g_{\mathrm{Mod}(t,\Delta t)}(x).
\end{equation}
where
\begin{equation}
    g_t(x)=\frac{e^{-x^2/4Dt}}{\sqrt{4\pi D t}}
\end{equation}
is the free Brownian propagator.

If the particle position is  measured uniformly in time, the corresponding distribution is stationary  
\begin{equation}
    P(x) = \frac{1}{\Delta t}\int_0^{\Delta t}\md t\,
    \frac{e^{-x^2/4Dt}}{\sqrt{4\pi D t}} .
\end{equation}
Performing the integral yields  
\begin{equation}\label{eq:BM_result_full}
    P(x) = 
        \frac{e^{-x^2/4D\Delta t}}{\sqrt{\pi D \Delta t}}
        - \frac{|x|}{2D \Delta t}  \,\mathrm{erfc}\!\left(\frac{|x|}{\sqrt{4D\Delta t}}\right),
\end{equation}
where $\mathrm{erfc}$ denotes the complementary error function.  

Turning now to the ABP, the heading angle $\theta$ diffuse and periodically reset to a prescribed angle $\psi$ (see Fig.~\ref{fig:setup}). In the fast-reset regime $\alpha\gg D$, the lateral spread of the particle position is small, and $\psi\simeq 0$. Therefore, the angle $\theta$ can be approximated to follow the Brownian resetting trajectory, leading to the stationary distribution 
\begin{equation}\label{eq:theta_distribution}
    P(\theta) = \sqrt{\frac{\alpha}{D}}
    \left[
        \frac{e^{-\alpha \theta^2 / 4D}}{\sqrt{\pi}}
        - \frac{|\theta|\sqrt{\alpha}}{\sqrt{4D}}
          \,\mathrm{erfc}\!\left(\frac{|\theta|\sqrt{\alpha}}{\sqrt{4D}}\right)
    \right],
\end{equation}
where we use $\Delta t =1/\alpha$. This gives the distribution $P(\theta)= \frac{1}{\sigma_\theta}f\left(\frac{\theta}{\sigma_\theta} \right)$ with standard deviation $\sigma_\theta=\sqrt{D/\alpha}$ and $f(x)$ in \eqref{eq:theta_distribution 1}.

\bibliography{resetting_bib}

@article{aguilar2018collective,
  title={Collective clog control: Optimizing traffic flow in confined biological and robophysical excavation},
  author={Aguilar, J and Monaenkova, D and Linevich, V and Savoie, W and Dutta, B and Kuan, H-S and Betterton, MD and Goodisman, MAD and Goldman, DI},
  journal={Science},
  volume={361},
  number={6403},
  pages={672--677},
  year={2018},
  publisher={American Association for the Advancement of Science}
}

@article{Majumdar2015Resetting,
  title = {Dynamical transition in the temporal relaxation of stochastic processes under resetting},
  author = {Majumdar, Satya N. and Sabhapandit, Sanjib and Schehr, Gr{\'e}gory},
  journal = {Physical Review E},
  volume = {91},
  number = {5},
  pages = {052131},
  year = {2015},
  publisher = {American Physical Society},
  doi = {10.1103/PhysRevE.91.052131}
}

@misc{yerrababu2024,
      title={Dynamical phase transitions in certain non-ergodic stochastic processes}, 
      author={Yogeesh Reddy Yerrababu and Satya N. Majumdar and Tridib Sadhu},
      year={2024},
      eprint={2412.19516},
      archivePrefix={arXiv},
      primaryClass={cond-mat.stat-mech},
      url={https://arxiv.org/abs/2412.19516}, 
}

@article{Romanczuk,
  title     = {Optimal chemotaxis in intermittent migration of animal cells},
  author    = {Romanczuk, P. and Salbreux, G.},
  journal   = {Physical Review E},
  volume    = {91},
  pages     = {042720},
  year      = {2015},
  publisher = {American Physical Society},
  doi       = {10.1103/PhysRevE.91.042720}
}

@article{ReichmanFried2004,
  title     = {Autonomous Modes of Behavior in Primordial Germ Cell Migration},
  author    = {Reichman-Fried, Michal and Minina, Sofia and Raz, Erez},
  journal   = {Developmental Cell},
  volume    = {6},
  pages     = {589--596},
  year      = {2004},
  publisher = {Cell Press},
  doi       = {10.1016/j.devcel.2004.03.003}
}

@article{Shavakh2022Zermelo,
  title     = {The generalization of Zermelo’s navigation problem with variable speed and limited acceleration},
  author    = {Mohammad Hossein Shavakh and Behroz Bidabad},
  journal   = {International Journal of Dynamics and Control},
  volume    = {10},
  number    = {2},
  pages     = {391--402},
  year      = {2022},
  doi       = {10.1007/s40435-021-00826-z},
  publisher = {Springer},
}

@article{bacik2025order,
  title={Order--disorder transition in multidirectional crowds},
  author={Bacik, Karol A and Sobota, Grzegorz and Bacik, Bogdan S and Rogers, Tim},
  journal={Proceedings of the National Academy of Sciences},
  volume={122},
  number={14},
  pages={e2420697122},
  year={2025},
  publisher={National Academy of Sciences}
}

@article{biswas2025drift,
  title={Drift-diffusive resetting search process with stochastic returns: Speedup beyond optimal instantaneous return},
  author={Biswas, Arup and Dubey, Ashutosh and Kundu, Anupam and Pal, Arnab},
  journal={Physical Review E},
  volume={111},
  number={1},
  pages={014142},
  year={2025},
  publisher={APS},
doi={https://doi.org/10.1103/PhysRevE.111.014142}
}

@article{tal2020experimental,
  title={Experimental realization of diffusion with stochastic resetting},
  author={Tal-Friedman, Ofir and Pal, Arnab and Sekhon, Amandeep and Reuveni, Shlomi and Roichman, Yael},
  journal={The journal of physical chemistry letters},
  volume={11},
  number={17},
  pages={7350--7355},
  year={2020},
  publisher={ACS Publications},
 doi={https://doi.org/10.1021/acs.jpclett.0c02122}
}

@article{BiroliMajumdarSchehr2025,
  author       = {Marco Biroli and Satya N. Majumdar and Gr\'egory Schehr},
  title        = {First Passage Resetting Gas},
  journal      = {arXiv preprint},
  eprint       = {2512.00440},
  eprinttype   = {arXiv},
  eprintclass  = {cond-mat.stat-mech},
  year         = {2025},
  note         = {Revised version (v2) submitted 11 Dec 2025},
  url          = {https://arxiv.org/abs/2512.00440}
}

@article{EvansMajumdar2019,
  author       = {Martin R. Evans and Satya N. Majumdar},
  title        = {Effects of refractory period on stochastic resetting},
  journal      = {Journal of Physics A: Mathematical and Theoretical},
  volume       = {52},
  number       = {1},
  pages        = {01LT01},
  year         = {2019},
  doi          = {10.1088/1751-8121/aaf080},
  eprint       = {arXiv:1809.01551},
  eprinttype   = {arXiv},
  eprintclass  = {cond-mat.stat-mech}
}

@article{mendez2024first,
  title={First-passage time of a Brownian searcher with stochastic resetting to random positions},
  author={Mendez, Vicen{\c{c}} and Flaquer-Galm{\'e}s, Rosa and Campos, Daniel},
  journal={Physical Review E},
  volume={109},
  number={4},
  pages={044134},
  year={2024},
  publisher={APS},
  doi={https://doi.org/10.1103/PhysRevE.109.044134}
}

@article{mori2025optimal,
  title={Optimal switching strategies for navigation in stochastic settings},
  author={Mori, Francesco and Mahadevan, L},
  journal={Journal of the Royal Society Interface},
  volume={22},
  number={227},
  pages={20240677},
  year={2025},
  publisher={The Royal Society},
  doi={https://doi.org/10.1098/rsif.2024.0677}
}

@article{besga2020optimal,
  title={Optimal mean first-passage time for a Brownian searcher subjected to resetting: experimental and theoretical results},
  author={Besga, Benjamin and Bovon, Alfred and Petrosyan, Artyom and Majumdar, Satya N and Ciliberto, Sergio},
  journal={Physical Review Research},
  volume={2},
  number={3},
  pages={032029},
  year={2020},
  publisher={APS},
doi={https://doi.org/10.1103/PhysRevResearch.2.032029}
}

@article{Romanczuk2015,  
title = {Optimal chemotaxis in intermittent migration of animal cells},  author = {Romanczuk, Pawel and Salbreux, Guillaume},  journal = {Physical Review E},  volume = {91},  number = {4},  pages = {042720},  year = {2015},  doi = {10.1103/PhysRevE.91.042720},  publisher = {American Physical Society}}

@article{Paramanick2024,  
title = {Uncovering Universal Characteristics of Homing Paths using Foraging Robots},  author = {Somnath Paramanick and Arup Biswas and Harsh Soni and Arnab Pal and Nitin Kumar},  journal = {PRX Life},  volume = {2},  pages = {033007},  year = {2024},  doi = {10.1103/PRXLife.2.033007},  publisher = {American Physical Society}}

@article{Baouche2025,  
title = {Optimal first-passage times of active Brownian particles under stochastic resetting},  author = {Yanis Baouche and Christina Kurzthaler},  journal = {Soft Matter},  volume = {21},  number = {29},  pages = {5998--6011},  year = {2025},  doi = {10.1039/d5sm00340g},  publisher = {Royal Society of Chemistry}}

@article{Shee2024,  title = {Active Brownian particle under stochastic position and orientation resetting in a harmonic trap},  author = {Amir Shee},  journal = {arXiv preprint arXiv:2409.06920},  year = {2024},  doi = {10.48550/arXiv.2409.06920},  url = {https://arxiv.org/abs/2409.06920}}

@article{Levine2023,  
title = {Physics of smart active matter: integrating active matter and control to gain insights into living systems},  author = {Herbert Levine and Daniel I. Goldman},  journal = {Soft Matter},  volume = {19},  number = {23},  pages = {4204--4207},  year = {2023},  doi = {10.1039/D3SM00171G},  publisher = {Royal Society of Chemistry}}

@article{Palagi2018Bioinspired,
  author    = {Stefano Palagi and Peer Fischer},
  title     = {Bioinspired microrobots},
  journal   = {Nature Reviews Materials},
  year      = {2018},
  volume    = {3},
  pages     = {113--124},
  doi       = {10.1038/s41578-018-0016-9},
  url       = {https://doi.org/10.1038/s41578-018-0016-9}
}

@article{Albers2002Exploring,
  author    = {Susanne Albers and Klaus Kursawe and Stefan Schuierer},
  title     = {Exploring Unknown Environments with Obstacles},
  journal   = {Algorithmica},
  year      = {2002},
  volume    = {32},
  number    = {1},
  pages     = {123--143},
  doi       = {10.1007/s00453-001-0067-x},
  url       = {https://doi.org/10.1007/s00453-001-0067-x}
}

@article{Naahidi2013Biocompatibility,
  author    = {Sheva Naahidi and Mousa Jafari and Faramarz Edalat and Kevin Raymond and Ali Khademhosseini and Pu Chen},
  title     = {Biocompatibility of engineered nanoparticles for drug delivery},
  journal   = {Journal of Controlled Release},
  year      = {2013},
  volume    = {166},
  number    = {2},
  pages     = {182--194},
  doi       = {10.1016/j.jconrel.2012.12.013},
  url       = {https://doi.org/10.1016/j.jconrel.2012.12.013}
}

@article{Doitsidou2002Cell,  author    = {Maria Doitsidou and Michal Reichman-Fried and J{\"u}rg Stebler and Marion K{\"o}prunner and Julia D{\"o}rries and Dirk Meyer and Camila V. Esguerra and Tin Chung Leung and Erez Raz},  title     = {Guidance of Primordial Germ Cell Migration by the Chemokine SDF-1},  journal   = {Cell},  year      = {2002},  volume    = {111},  number    = {5},  pages     = {647--659},  doi       = {10.1016/S0092-8674(02)01135-2},  url       = {https://doi.org/10.1016/S0092-8674(02)01135-2}}

@article{Endres2008PNAS,
  author    = {Robert G. Endres and Ned S. Wingreen},
  title     = {Accuracy of direct gradient sensing by single cells},
  journal   = {Proceedings of the National Academy of Sciences of the United States of America},
  year      = {2008},
  volume    = {105},
  number    = {41},
  pages     = {15749--15754},
  doi       = {10.1073/pnas.0804686105},
  url       = {https://doi.org/10.1073/pnas.0804686105}
}

@article{Levine2013PhysicsToday,  author    = {Herbert Levine and Wouter-Jan Rappel},  title     = {The physics of eukaryotic chemotaxis},  journal   = {Physics Today},  year      = {2013},  volume    = {66},  number    = {2},  pages     = {24--30},  doi       = {10.1063/PT.3.1884},  url       = {https://doi.org/10.1063/PT.3.1884}}

@article{Swaney2010Eukaryotic,
  author    = {Kristen F. Swaney and Chuan-Hsiang Huang and Peter N. Devreotes},
  title     = {Eukaryotic Chemotaxis: A Network of Signaling Pathways Controls Motility, Directional Sensing, and Polarity},
  journal   = {Annual Review of Biophysics},
  year      = {2010},
  volume    = {39},
  pages     = {265--289},
  doi       = {10.1146/annurev.biophys.093008.131228},
  url       = {https://doi.org/10.1146/annurev.biophys.093008.131228}
}

@article{Reynolds2007,
  title   = {Lévy Flight Patterns of Wandering Albatrosses},
  author  = {Andy M. Reynolds and David W. Sims and Graeme M. Viswanathan and others},
  journal = {Nature},
  volume  = {450},
  pages   = {77--80},
  year    = {2007},
  doi     = {10.1038/nature06199},
  publisher = {Nature Publishing Group}
}

@book{Stone1989,  title     = {Theory of Optimal Search},  author    = {Lawrence D. Stone},  edition   = {2nd},  year      = {1989},  publisher = {Military Applications Section, Operations Research Society of America},  address   = {Arlington, VA},  isbn      = {187764000X},  pages     = {279}}

@article{Bartumeus2002,  title   = {Optimizing the Encounter Rate in Biological Interactions: Lévy versus Brownian Strategies},  author  = {F. Bartumeus and J. Catalan and U. L. Fulco and M. L. Lyra and G. M. Viswanathan},  journal = {Physical Review Letters},  volume  = {88},  pages   = {097901},  year    = {2002},  doi     = {10.1103/PhysRevLett.88.097901},  publisher = {American Physical Society}}

@article{Eisenbach2006,
  title   = {Sperm guidance in mammals — an unpaved road to the egg},
  author  = {Michael Eisenbach and Laura C. Giojalas},
  journal = {Nature Reviews Molecular Cell Biology},
  volume  = {7},
  number  = {4},
  pages   = {276--285},
  year    = {2006},
  doi     = {10.1038/nrm1893},
  publisher = {Nature Publishing Group}
}

@article{Meerson2015,
  title = {Mortality, Redundancy, and Diversity in Stochastic Search},
  author = {Baruch Meerson and S. Redner},
  journal = {Physical Review Letters},
  volume = {114},
  pages = {198101},
  year = {2015},
  doi = {10.1103/PhysRevLett.114.198101},
  publisher = {American Physical Society}
}

@incollection{Basu2024,  
author    = {Urna Basu and Sanjib Sabhapandit and Ion Santra},  title     = {Target search by active particles},  booktitle = {Target Search Problems},  editor    = {D. S. Grebenkov and R. Metzler and G. Oshanin},  publisher = {Springer Nature Switzerland},  year      = {2024},  pages     = {463--487},  doi       = {10.1007/978-3-031-67802-8_19},  url       = {https://link.springer.com/chapter/10.1007/978-3-031-67802-8_19}}

@article{Casert2024,  
title = {Learning protocols for the fast and efficient control of active matter},  author = {Casert, Corneel and Whitelam, Stephen},  journal = {Nature Communications},  volume = {15},  number = {1},  pages = {52878},  year = {2024},  doi = {10.1038/s41467-024-52878-2},  publisher = {Nature Publishing Group}}

@article{Kumar2020,  
title = {Active Brownian motion in two dimensions under stochastic resetting},  author = {Vijay Kumar and Onkar Sadekar and Urna Basu},  journal = {Physical Review E},  volume = {102},  pages = {052129},  year = {2020},  doi = {10.1103/PhysRevE.102.052129},  publisher = {American Physical Society}}

@article{Altshuler2024,  
title = {Environmental memory facilitates search with home returns},  author = {Amy Altshuler and Ofek Lauber Bonomo and Nicole Gorohovsky and Shany Marchini and Eran Rosen and Ofir Tal-Friedman and Shlomi Reuveni and Yael Roichman},  journal = {Physical Review Research},  volume = {6},  pages = {023255},  year = {2024},  doi = {10.1103/PhysRevResearch.6.023255},  publisher = {American Physical Society}}

@INPROCEEDINGS{11268231,
  author={Samolej, Slawomir and Baturyna, Darya and Davidrajuh, Reggie},
  booktitle={2025 9th International Symposium on Multidisciplinary Studies and Innovative Technologies (ISMSIT)}, 
  title={Warehouse Automation Simulation involving Robotic Picking System}, 
  year={2025},
  volume={},
  number={},
  pages={1-6},
  doi={10.1109/ISMSIT67332.2025.11268231}}

@article{evans2020stochastic,
  title={Stochastic resetting and applications},
  author={Evans, M. R. and Majumdar, S. N. and Schehr, G.},
  journal={Journal of Physics A: Mathematical and Theoretical},
  volume={53},
  number={19},
  pages={193001},
  year={2020},
  doi={10.1088/1751-8121/ab7cfe}
}

@article{bechinger2016active,
  title={Active particles in complex and crowded environments},
  author={Bechinger, Clemens and Di Leonardo, Roberto and L{\"o}wen, Hartmut and Reichhardt, Charles and Volpe, Giorgio and Volpe, Giovanni},
  journal={Reviews of modern physics},
  volume={88},
  number={4},
  pages={045006},
  year={2016},
  publisher={APS},
  doi={10.1103/RevModPhys.88.045006}

}

@article{cates2015motility,
  title={Motility-induced phase separation},
  author={Cates, Michael E and Tailleur, Julien},
  journal={Annu. Rev. Condens. Matter Phys.},
  volume={6},
  number={1},
  pages={219--244},
  year={2015},
  publisher={Annual Reviews},
  doi={10.1146/annurev-conmatphys-031214-014710}
}

@article{santra2020run,
  title={Run-and-tumble particles in two dimensions under stochastic resetting conditions},
  author={Santra, Ion and Basu, Urna and Sabhapandit, Sanjib},
  journal={Journal of Statistical Mechanics: Theory and Experiment},
  volume={2020},
  number={11},
  pages={113206},
  year={2020},
  publisher={IOP Publishing}
}

@article{benichou2011intermittent,
  title={Intermittent search strategies},
  author={B{\'e}nichou, Olivier and Loverdo, Claude and Moreau, Michel and Voituriez, Raphael},
  journal={Reviews of Modern Physics},
  volume={83},
  number={1},
  pages={81--129},
  year={2011},
  publisher={APS},
  doi={10.1103/RevModPhys.83.81}
}

@article{benichou2008optimizing,
  title={Optimizing intermittent reaction paths},
  author={B{\'e}nichou, O and Loverdo, C and Moreau, M and Voituriez, R},
  journal={Physical Chemistry Chemical Physics},
  volume={10},
  number={47},
  pages={7059--7072},
  year={2008},
  publisher={Royal Society of Chemistry}
}

@article{thrun2002probabilistic,
  title={Probabilistic robotics},
  author={Thrun, Sebastian},
  journal={Communications of the ACM},
  volume={45},
  number={3},
  pages={52--57},
  year={2002},
  publisher={ACM New York, NY, USA}
}

@inproceedings{wang2016apriltag,
  title={AprilTag 2: Efficient and robust fiducial detection},
  author={Wang, John and Olson, Edwin},
  booktitle={2016 IEEE/RSJ International Conference on Intelligent Robots and Systems (IROS)},
  pages={4193--4198},
  year={2016},
  organization={IEEE},
  doi={10.1109/IROS.2016.7759617}
}

@book{bertsekas2021data,
  title={Data networks},
  author={Bertsekas, Dimitri and Gallager, Robert},
  year={2021},
  publisher={Athena Scientific}
}

@book{siegwart2011introduction,
  title={Introduction to autonomous mobile robots},
  author={Siegwart, Roland and Nourbakhsh, Illah Reza and Scaramuzza, Davide},
  year={2011},
  publisher={MIT press}
}

@article{howse2007self,
  title={Self-motile colloidal particles: from directed propulsion to random walk},
  author={Howse, Jonathan R and Jones, Richard AL and Ryan, Anthony J and Gough, Tim and Vafabakhsh, Reza and Golestanian, Ramin},
  journal={Physical review letters},
  volume={99},
  number={4},
  pages={048102},
  year={2007},
  publisher={APS},
  doi={10.1103/PhysRevLett.99.048102}
}

@article{fily2012athermal,
  title={Athermal phase separation of self-propelled particles with no alignment},
  author={Fily, Yaouen and Marchetti, M Cristina},
  journal={Physical review letters},
  volume={108},
  number={23},
  pages={235702},
  year={2012},
  publisher={APS},
  doi={10.1103/PhysRevLett.108.235702}
}

@article{INA219,
  title={A high-side current shunt and power monitor with I$^2$C interface for embedded systems},
  author={{Texas Instruments}},
  journal={Datasheet INA219},
  year={2019},
  note={Available at: \url{https://www.ti.com/lit/ds/symlink/ina219.pdf}}
}

@article{corbetta2017fluctuations,
  title={Fluctuations around mean walking behaviors in diluted pedestrian flows},
  author={Corbetta, Alessandro and Lee, Chung-min and Benzi, Roberto and Muntean, Adrian and Toschi, Federico},
  journal={Physical Review E},
  volume={95},
  number={3},
  pages={032316},
  year={2017},
  publisher={APS},
  doi={10.1103/PhysRevE.95.032316}
}

@article{schwarting2018planning,
  title={Planning and decision-making for autonomous vehicles},
  author={Schwarting, Wilko and Alonso-Mora, Javier and Rus, Daniela},
  journal={Annual Review of Control, Robotics, and Autonomous Systems},
  volume={1},
  number={1},
  pages={187--210},
  year={2018},
  publisher={Annual Reviews},
  doi={10.1146/annurev-control-060117-105157}
}

@article{evans2011diffusion,
  title={Diffusion with stochastic resetting},
  author={Evans, Martin R and Majumdar, Satya N},
  journal={Physical review letters},
  volume={106},
  number={16},
  pages={160601},
  year={2011},
  publisher={APS}
}

@article{fuchs2016stochastic,
  title={Stochastic thermodynamics of resetting},
  author={Fuchs, Jaco and Goldt, Sebastian and Seifert, Udo},
  journal={Europhysics Letters},
  volume={113},
  number={6},
  pages={60009},
  year={2016},
  publisher={IOP Publishing}
}

@article{Hays2020OpenOceanReorientation,
  title        = {Open Ocean Reorientation and Challenges of Island Finding by Sea Turtles during Long-Distance Migration},
  author       = {Hays, Graeme C. and Cerritelli, Giulia and Esteban, Nicole and Rattray, Alex and Luschi, Paolo},
  journal      = {Current Biology},
  year         = {2020},
  volume       = {30},
  number       = {14},
  pages        = {R835--R836},
  doi          = {10.1016/j.cub.2020.05.08},
  publisher    = {Cell Press}
}

@article{AuthorLastname2025Title,
  title        = {Route optimisation and solving Zermelo’s navigation problem during long‐distance migration in cross flows},
  author       = {Hays, Graeme C. and Cerritelli, Giulia and Esteban, Nicole and Rattray, Alex and Luschi, Paolo},
  journal      = {Ecology Letters},
  year         = {2025},
  doi          = {10.1111/ele.12219}
}

@article{Hays2022TravelRoutesToRemote,
  author       = {Hays, Graeme C. and Laloë, Julian-Olivier and Rattray, Alistair and Esteban, Nicole},
  title        = {Travel routes to remote ocean targets reveal the map sense resolution for a marine migrant},
  journal      = {Journal of the Royal Society Interface},
  year         = {2022},
  volume       = {19},
  number       = {192},
  pages        = {20210859},
  doi          = {10.1098/rsif.2021.0859},
  url          = {https://doi.org/10.1098/rsif.2021.0859}
}

@article{NaisbettJones2017MagneticMapEels,
  author       = {Naisbett-Jones, L. C. and Stokesbury, M. J. W. and Limburg, K. E. and Putman, N. F.},
  title        = {A Magnetic Map Leads Juvenile European Eels to the Gulf of Mexico},
  journal      = {Current Biology},
  year         = {2017},
  volume       = {27},
  number       = {23},
  pages        = {R1266--R1267},
  doi          = {10.1016/j.cub.2017.09.045},
  url          = {https://doi.org/10.1016/j.cub.2017.09.045}
}

@article{Thorup2011JuvenileSongbirds,
  title        = {Juvenile Songbirds Compensate for Displacement to Oceanic Islands during Autumn Migration},
  author       = {Thorup, Kasper and Ortvad, Troels Eske and Rab{\o}l, J{\o}rgen and Holland, Richard A. and T{\o}ttrup, Anders P. and Wikelski, Martin},
  journal      = {PLOS ONE},
  year         = {2011},
  volume       = {6},
  number       = {3},
  pages        = {e17903},
  doi          = {10.1371/journal.pone.0017903},
  url          = {https://doi.org/10.1371/journal.pone.0017903}
}

@article{Horton2016Nocturnally,
  author       = {Horton, Kyle G. and Van Doren, Benjamin M. and Stepanian, Phillip M. and Hochachka, Wesley M. and Farnsworth, Andrew and Kelly, Jeffrey F.},
  title        = {Nocturnally migrating songbirds drift when they can and compensate when they must},
  journal      = {Scientific Reports},
  year         = {2016},
  volume       = {6},
  pages        = {21249},
  doi          = {10.1038/srep21249},
  url          = {https://doi.org/10.1038/srep21249}
}

@article{McLaren2022Predicting,
  author       = {James D. McLaren and Heiko Schmaljohann and Bernd Blasius},
  title        = {Predicting performance of naïve migratory animals, from many wrongs to self-correction},
  journal      = {Communications Biology},
  volume       = {5},
  number       = {1},
  pages        = {1058},
  year         = {2022},
  doi          = {10.1038/s42003-022-03995-5},
  url          = {https://doi.org/10.1038/s42003-022-03995-5}
}

@article{Merlin2012NavigationalStrategies,
  title        = {Unraveling navigational strategies in migratory insects},
  author       = {Merlin, Christine and Heinze, Stanley and Reppert, Steven M.},
  journal      = {Current Opinion in Neurobiology},
  year         = {2012},
  volume       = {22},
  number       = {2},
  pages        = {353--361},
  doi          = {10.1016/j.conb.2011.11.009},
  url          = {https://doi.org/10.1016/j.conb.2011.11.009}
}

@article{Liebchen2019Optimal,
  author       = {Benno Liebchen and Hartmut L\"owen},
  title        = {Optimal navigation strategies for active particles},
  journal      = {EPL (Europhysics Letters)},
  volume       = {127},
  number       = {3},
  pages        = {34003},
  year         = {2019},
  doi          = {10.1209/0295-5075/127/34003},
  url          = {https://doi.org/10.1209/0295-5075/127/34003}
}

@article{Fernandez-Rodriguez2020,
  author       = {Miguel Angel Fernandez-Rodriguez and Fabio Grillo and Laura Alvarez and Marco Rathlef and Ivo Buttinoni and Giovanni Volpe and Lucio Isa},
  title        = {Feedback-controlled active Brownian colloids with space-dependent rotational dynamics},
  journal      = {Nature Communications},
  volume       = {11},
  number       = {4223},
  year         = {2020},
  doi          = {10.1038/s41467-020-17864-4},
  url          = {https://www.nature.com/articles/s41467-020-17864-4}
}

@article{DeBruyneMori2023Resetting,
  author       = {Benjamin De Bruyne and Francesco Mori},
  title        = {Resetting in Stochastic Optimal Control},
  journal      = {Physical Review Research},
  volume       = {5},
  number       = {1},
  pages        = {013122},
  year         = {2023},
  doi          = {10.1103/PhysRevResearch.5.013122},
  url          = {https://doi.org/10.1103/PhysRevResearch.5.013122}
}

@article{Putzke2023,
  author       = {Mischa Putzke and Holger Stark},
  title        = {Optimal navigation of a smart active particle: directional and distance sensing},
  journal      = {The European Physical Journal E},
  volume       = {46},
  number       = {6},
  pages        = {48},
  year         = {2023},
  doi          = {10.1140/epje/s10189-023-00309-3},
  url          = {https://doi.org/10.1140/epje/s10189-023-00309-3}
}

\end{document}


\setcounter{secnumdepth}{3}

\title{Supplementary Information for the article ``Optimal navigation in a noisy environment"}
\author{Abhijit Sinha} 
 \affiliation{Department of Condensed Matter Physics and Materials Science, Tata Institute of Fundamental Research, Homi Bhabha Road, Mumbai 400005, India.}
 \author{Sandeep Jangid}
\affiliation{Department of Theoretical Physics, Tata Institute of Fundamental Research, Homi Bhabha Road, Mumbai 400005, India.}
 \author{Tridib Sadhu}
\affiliation{Department of Theoretical Physics, Tata Institute of Fundamental Research, Homi Bhabha Road, Mumbai 400005, India.}
\author{Shankar Ghosh} 
 \affiliation{Department of Condensed Matter Physics and Materials Science, Tata Institute of Fundamental Research, Homi Bhabha Road, Mumbai 400005, India.}
\begin{abstract}
This Supplementary Information presents additional experimental details supporting the main article. We describe the robotic platform, experimental arena, and real-time control framework, along with calibration and power consumption measurements. The stochastic heading dynamics are characterized through measurements of the angular diffusion constant, resetting time, and transverse mean-square displacement. We also discuss sources of locomotion uncertainty and analyze head-on collision statistics under periodic resetting.
\end{abstract}
\maketitle
\clearpage

\section{Robot Setup}

\paragraph{\textbf{Construction of the robot:}}

\begin{figure}
    \centering
    \includegraphics[width=0.95\linewidth]{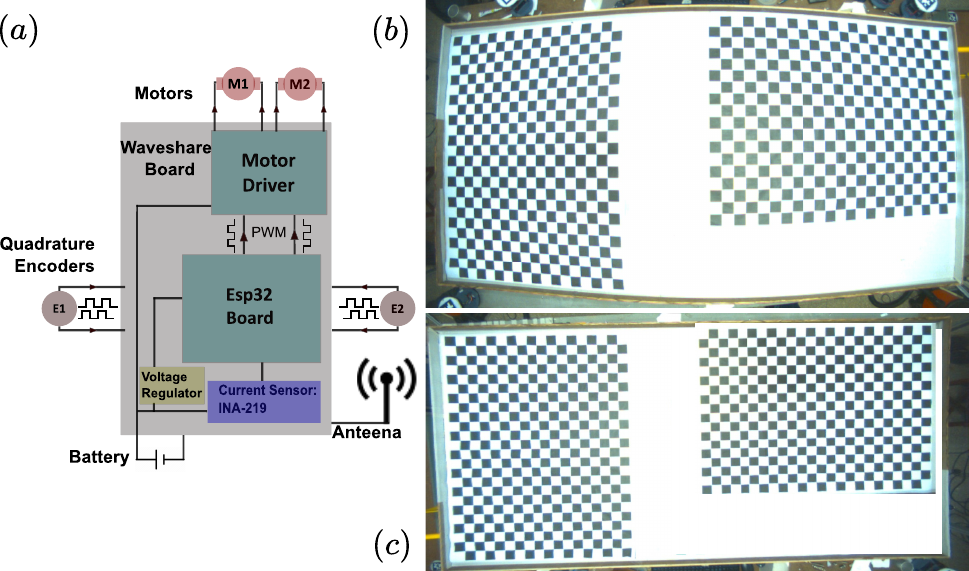}
    \caption{\textbf{Robot Setup:} (a) Differential Driven Robot Setup. (b), (c) Fish eye correction before and after the calibration. }
    \label{fig:calibration}
\end{figure}

As demonstrated, each Fig. \ref{fig:calibration} (a) robot is a differential-drive platform built on a compact 3D-printed chasis. Locomotion is provided by two N20-type 3 V geared DC motors, each equipped with an integrated Hall-effect quadrature encoder. The quadrature signals supply two phase-shifted pulse trains, allowing us to measure wheel rotation with direction sensitivity. A pair of passive caster wheels supports the chassis, ensuring stable planar motion.
The robot's motion controller consists of an integrated Wave share driver board powered by
three 3.7~V lithium-ion cells in series (11.1~V nominal). The board consists of an integrated ESP32 microcontroller, which has a Tensilica Xtensa dual-core 32-bit LX6 microprocessor. The board also includes H-bridge motor driver (TB6612FNG)
, which is capable of controlling two motors and encoder interfaces,
allowing wheel velocities to be regulated through PWM.

\paragraph{\textbf{The Experimental Arena:}}
The experiments are conducted in a wooden robotic arena measuring
$50\times100$~inches. A Pixielink camera equipped with a fisheye lens is
mounted above the arena to track the AprilTags attached to the top of the robot. Image
acquisition is performed through a dedicated camera handler that interfaces
directly with the hardware. The camera is geometrically calibrated using a
$48\times38$~inch checkerboard; a calibration routine collects several
$800\times1400$ pixel images at different orientations and determines the
intrinsic parameters and distortion coefficients (see Fig.\ref{fig:calibration}(b),(c) used for all subsequent
position reconstruction.

\paragraph{\textbf{The Control Software:}}
The control software follows a modular, multi-threaded architecture in which
each functional component operates independently. The main control thread
retrieves frames from the camera handler, detects AprilTags, updates the robot
state, and issues commands to the robots. AprilTag detection is the dominant
computational step: processing a full $800\times1400$ frame requires
approximately $20\,\mathrm{ms}$. To maintain real-time operation, only the
first frame is processed in full; subsequent detections are restricted to a
small region-of-interest centered on the robot's previously estimated
location, reducing the detection time to $\sim2\,\mathrm{ms}$. Distortion
correction is applied only to detected tag coordinates and it does not
significantly contribute to latency. Communication with each robot is handled by an independent Client
thread, which maintains a TCP connection, receives status messages, and
executes commands supplied by the controller. The robot transmits a
``STOPPED'' acknowledgement once a command is completed, ensuring synchronization
between state estimation and actuation. Logging is performed in parallel
through two additional modules: a tag logger that records detection data, and
an logger that stores the robot’s communication messages. A separate data
saver thread stores detected, robot's centroid and orientations, in text files
without interfering with real-time control. By decoupling the camera handler, AprilTag detector, controller, Client,
logging modules, and data saver into independent threads, the architecture
prevents blocking between subsystems and ensures stable real-time performance
even with high-resolution images and multiple concurrent tasks. Power consumption is monitored using an INA219 sensor, which reports voltage, current, and power over an I$^2$C interface. ESP32 microcontroller running a lightweight, parallelized firmware that combines motor actuation, quadrature-encoder odometry, energy sensing, and WiFi communication. Hall-effect encoder signals are captured via hardware interrupts to obtain wheel-rotation counts, which are converted to RPM using the known gearbox ratio. The control logic is divided into two FreeRTOS tasks: one receives WiFi commands (forward, backward, turning, stop, and a short randomized “wiggle’’ routine), resets encoder counters, and executes the requested motion for a specified duration while the second task reports timing, RPMs, and power draw whenever the motors stop. The wiggle routine consists of brief alternating motor pulses at $\sim$150~ms intervals, producing small random reorientation kicks. A simple TCP link handles acknowledgments and feedback, and automatic reconnection ensures robustness to transient WiFi loss. This parallel, nonblocking structure provides a responsive and low-latency control loop suitable for multi-robot experiments.

\paragraph{\textbf{Power Consumption in the different modes:}}
\begin{figure}
       \centering
       \includegraphics[width=0.95\linewidth]{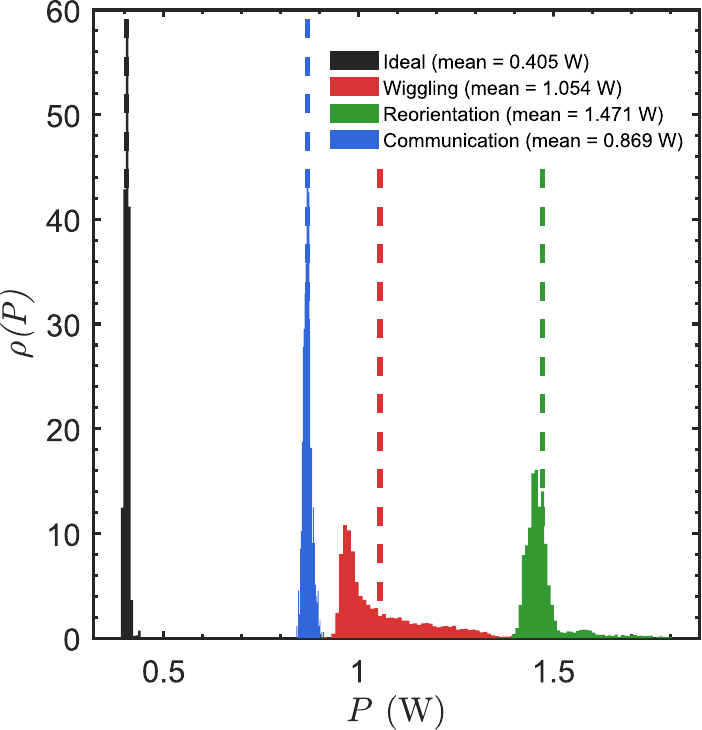}
       \caption{\textbf{$\rho(P)$ vs $P$ (W)}: Probability density of power for different robot states—Ideal, Wiggling, Reorienting toward the destination, and Communicating.The mean power values for each state are indicated by dotted lines, with their numerical values shown in the legend.}

     \label{fig:Power}
\end{figure}

Here we  describe the experimental setup and methodology used to measure the robot’s power consumption across different operational modes and to calibrate its onboard current sensor. The experiment was designed to provide accurate, high-resolution electrical measurements for developing a detailed energy consumption profile of the system. The robot operates in four distinct modes, either individually or in combination: Idle, where the system is powered but inactive; Wireless Communication, during which the robot transmits or receives data over a wireless link; Wiggling, an oscillatory motion used for forward movement; and Turning in Place, a maneuver in which the wheels rotate in opposite directions to adjust the robot's heading without translational motion. To perform precise voltage and current measurements, two digital multimeters were used: a Keithley 2000, configured as a DC voltmeter and connected in parallel with the battery terminals, and a Keithley 2000, configured as a DC ammeter and connected in series with the Waveshare 23730-based control board. Power was supplied by a standard 3,  18650-Li-ion battery. Both instruments were interfaced with a PC via a GPIB-USB-HS adapter from National Instruments. Data acquisition and control were handled using MATLAB, utilizing the Instrument Control Toolbox and NI-VISA drivers for GPIB communication. MATLAB scripts were developed to initialize and configure the instruments and to record measurements at a sampling rate of 10 Hz, which provided sufficient temporal resolution to capture power variations across all modes. The robot was programmed to cycle through each mode, with each mode lasting 5 seconds. This entire sequence was repeated 100 times. Instantaneous power $P(t)$ was computed  from the  values of voltage $V(t)$  and current $I(t)$, respectively. This setup enabled the collection of consistent and reliable power data for each operating mode and served as the basis for calibrating the robot's onboard current sensor ( INA219). The  histograms of the power drawn in each mode is shown in Fig. \ref{fig:Power}.

\section{Mean Square Displacement (MSD) Analysis and Energy-Aware Navigation}

\subsection{Transverse Mean Square Displacement Dynamics}

The impact of the periodic heading resets on the robot's motion is quantified by analyzing the \textbf{transverse mean-squared displacement (MSD)}, $\text{MSD}(\tau ) = \langle [\Delta y(t+\tau) - \Delta y(t)]^2 \rangle$, as a function of the lag time $\Delta t$.

\begin{figure}
    \centering
    \includegraphics[width=0.95\linewidth]{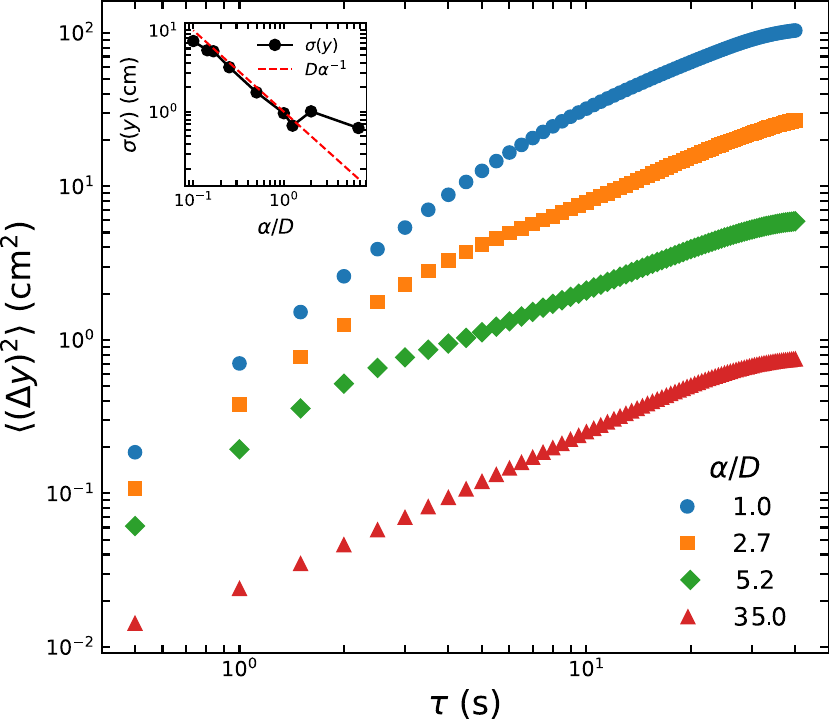}
    \caption{Mean square displacement (MSD, in cm$^2$) of the robot in the transverse ($\Delta y$) direction as a function of time lag $\tau$ (in seconds) for different normalized resetting rates $\alpha/D$. The inset shows the standard deviation $\sigma(\Delta y)$ versus $\alpha/D$, with the red dotted line representing the reference curve $\sigma(\Delta y) \sim D /\alpha$.}
    \label{fig:MSD}
\end{figure}

Figure~\ref{fig:MSD} shows the transverse MSD for various normalized resetting rates $\alpha/D$. For early  lag times  the MSD exhibits ballistic growth ($\sim \tau^2$). This regime corresponds to the persistent, free motion of the robot between consecutive directional resets. Beyond a characteristic timescale the MSD growth significantly slows down, trending towards saturation at a plateau value. This trend toward saturation signifies the effective spatial confinement introduced by the frequent reorientations. The resets successfully suppress large-scale, uncorrected directional drift, keeping the trajectory localized around the mean path.
As illustrated in Figure~\ref{fig:MSD}, an increase in the reset rate, $\alpha$, actively suppresses large transverse excursions, effectively confining the robot's trajectory to a narrow spatial band around the average path. Conversely, when resets are infrequent (low $\alpha$), the absence of timely directional correction allows for persistent random walks, resulting in significantly broader spatial wandering.

This confinement is quantitatively captured by the transverse positional standard deviation, $\sigma(\Delta y)$, which is shown in the inset of Figure~\ref{fig:MSD} as a function of $\alpha/D$. For lower reset rates, the data aligns with the expected scaling for diffusive confinement under periodic resetting, $\sigma(\Delta y) \sim D/\alpha$. However, at high $\alpha$, $\sigma(\Delta y)$ is observed to saturate. This plateau is attributed to the intrinsic heading tolerance (approximately $\pm5^{\circ}$) of the low-cost robotic platform, which imposes a physical limit on the achievable alignment precision, regardless of the reset frequency.

\section{Measurement of the Angular Diffusion Constant (\textit{D})}

This section details the mechanism used for the robot's forward movement and derives the effective angular diffusion constant ($D$) characterizing the resulting stochastic heading dynamics.

The robot’s forward propulsion is achieved through a ``wiggle mode'' that implements a minimal stochastic control policy. At any instant $t$, the differential-drive robot operates with wheel angular velocities $(\omega_L,\omega_R)$. In wiggle mode, the robot executes only two elementary actions at fixed time intervals $\epsilon$: left-wheel drive $(\omega_0,0)$ or right-wheel drive $(0,\omega_0)$, selected randomly with equal probability $1/2$. This random steering produces small, unbiased angular increments $\delta\gamma_i$ in the robot’s heading, resulting in diffusive orientation dynamics.

Assuming pure rolling, the linear wheel speed is $v_0=\omega_0 R$, where $R$ is the wheel radius.
For a single elementary step of duration $\epsilon$ $=0.2$ sec, the incremental change in heading $\delta\gamma_0$ follows from the geometric relation:
\begin{equation}
    \delta\gamma_0\,\frac{L_{\mathrm{axel}}}{2}
    = v_0\,\epsilon
    = R\omega_0\,\epsilon
\end{equation}
where $L_{\mathrm{axel}}$ is the distance between the wheel axes.
Rearranging this equation yields the magnitude of the angular increment:
$$\delta\gamma_0 = \frac{2 R \omega_0 \epsilon}{L_{\mathrm{axel}}}$$

After $n$ such steps, the total time elapsed is $t=n\epsilon$, and the total accumulated heading change is $\gamma=\sum_{i=1}^n\delta\gamma_i$. Since the steering direction is random and unbiased, the mean accumulated heading change is zero, $\langle\gamma\rangle=0$.

The \textbf{mean-squared angular displacement} grows linearly with time ($t$):
\begin{equation}
    \langle\gamma^2\rangle
    = n\,\delta\gamma_0^2
    = \frac{t}{\epsilon} \left(\frac{2 R \omega_0 \epsilon}{L_{\mathrm{axel}}}\right)^2
    = 4t\epsilon\left(\frac{R\omega_0}{L_{\mathrm{axel}}}\right)^2
\end{equation}
The effective \textbf{angular diffusion constant $D$} is defined through the relation $\langle\gamma^2\rangle=2Dt$. Solving for $D$:
\begin{equation}
    D = \frac{\langle\gamma^2\rangle}{2t}
    = 2\epsilon\left(\frac{R\omega_0}{L_{\mathrm{axel}}}\right)^2
\end{equation}
Using the experimentally measured geometric and actuation parameters specific to the robot, we obtain the calculated value:
$D = 0.09$

\begin{figure}
    \centering
    \includegraphics[width=0.95\linewidth]{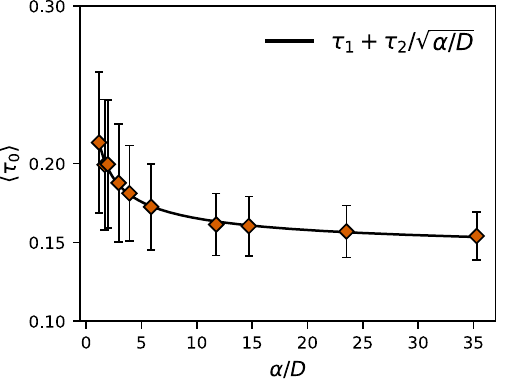}
    \caption{Mean Time to Reset $\langle\tau_{0}\rangle$ versus normalized resetting rate $\alpha/D$. The figure illustrates the time required for the robot to complete a full heading reset maneuver. The experimental data fits  to the functional dependency:
        $\langle \tau_0 \rangle = \tau_1 + \frac{\tau_2}{\sqrt{\alpha/D}}$
        The constant term $\tau_0$ represents the intrinsic time overhead incurred by the physical reorientation mechanism, while the second term accounts for the time spent correcting the angular drift, governed by the characteristic diffusive timescale $\tau_2$.}
    \label{fig:Reset_time}
\end{figure}

\section{Measurement of the Resetting time as a  function of the resetting rate alpha}

The    Fig. \ref{fig:Reset_time} shows that that the  Mean Time to reset $\langle\tau_{0}\rangle$   decreases as a function of the  normalised resetting time $\alpha/D$.
The experimentally measured mean characteristic resetting time
$\langle\tau_0\rangle$ is well described by the scaling form
\begin{equation}
    \langle\tau_0\rangle
    = \tau_1 + \frac{\tau_2}{\sqrt{\alpha/D}} ,
\end{equation}
where $\tau_1$ represents a constant overhead associated with reorientation, and
$\tau_2$ is a characteristic timescale set by diffusive angular wandering. The
nonuniform dependence of $\langle\tau_0\rangle$ on the normalized resetting rate
$\alpha /D$ is discussed in the main text.

 \section{Uncertainties in Locomotion and Heading Resetting \label{app:robot characteristics}}
\begin{figure*}
\centering
\includegraphics[width=0.95\linewidth]{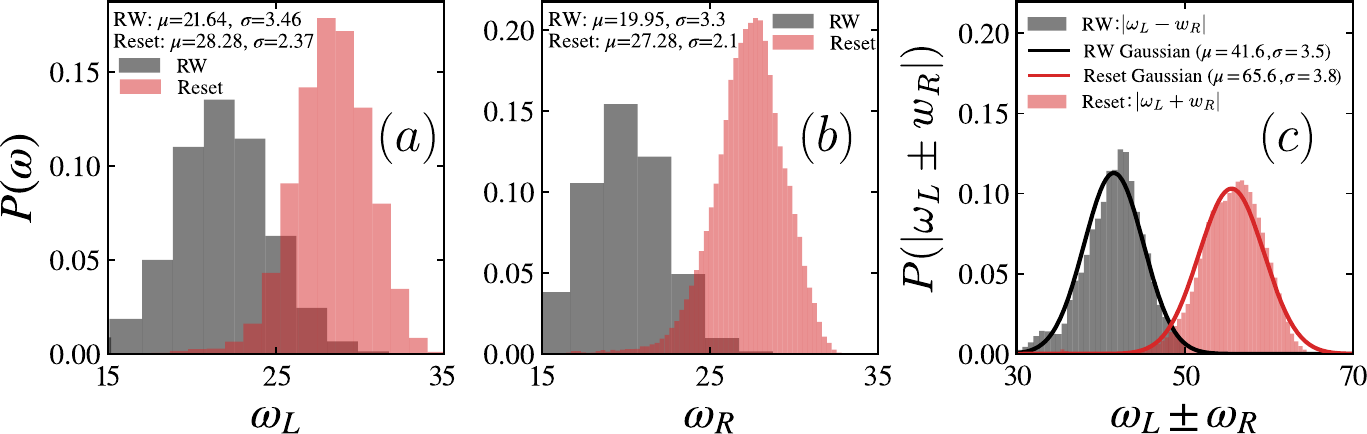}
 \caption{(a), (b) Speed distributions of left and right motors measured from encoders. (c) Distribution of $\omega_{L}\pm\omega_{R}$.
These distributions highlight variability in actuation, one of the primary contributors to process noise.}
\label{fig:encoder}
\end{figure*}

The purpose of this section is to identify the main sources of error that arise
in our experiments and to describe how they enter the robot’s dynamics.

At any instant the pose of the robot is represented as
\[
\mathbf{x}_t = [\,x_t,\; y_t,\; \theta_t\,],
\]
with $(x_t,y_t)$ denoting planar coordinates and $\theta_t$ the heading.
The control inputs are derived from wheel encoders,
\[
v_t = \tfrac{r}{2}(\omega_R+\omega_L), \qquad
\omega_t = \tfrac{r}{L_{\mathrm{axel}}}(\omega_R-\omega_L),
\]
where $r = 17$mm is the wheel radius, $L_{\mathrm{axel}}=75$mm the axle length, and $\omega_{L,R}$ the
wheel angular velocities. Thus $\mathbf{u}_t = [\,v_t,\;\omega_t\,]$. Figure~\ref{fig:encoder} shows representative experimentally measured probability distributions of the left and right wheel angular velocities, $\omega_L$ and $\omega_R$, as well as of the derived quantities $|\omega_L \pm \omega_R|$.

Between resets, the pose evolves according to odometry,
\[
\mathbf{x}_{t+1} = f(\mathbf{x}_t,\mathbf{u}_t) + Q_t,
\]
where $Q_t$ represents process noise. This noise reflects a combination of
encoder quantization, wheel slip, surface irregularities, and the deliberately
wiggly trajectory employed to cancel systematic left–right bias. While this
strategy reduces systematic drift, it introduces additional random fluctuations,
which accumulate as the robot advances.

At reset events, the robot pose is corrected using AprilTag detections,
\[
\mathbf{z}_t = \mathbf{x}_t + R_t,
\]
where $R_t$ denotes measurement noise. This uncertainty arises from the finite
accuracy of camera-based position estimates as well as the limited precision of
heading realignment. In practice, the robot’s orientation can only be reset to
within $\pm 5^\circ$ of the target, so the residual misalignment is absorbed
into $R_t$.

In summary, the variability in the robot’s motion originates from two distinct
sources. Process noise accumulates during odometry because of actuation
uncertainties and intentional wiggling, while measurement noise enters through
the finite precision of vision-based resets.

\section{Head-on Collisions Between Robots}

\begin{figure*}
\includegraphics[width=0.95\linewidth]{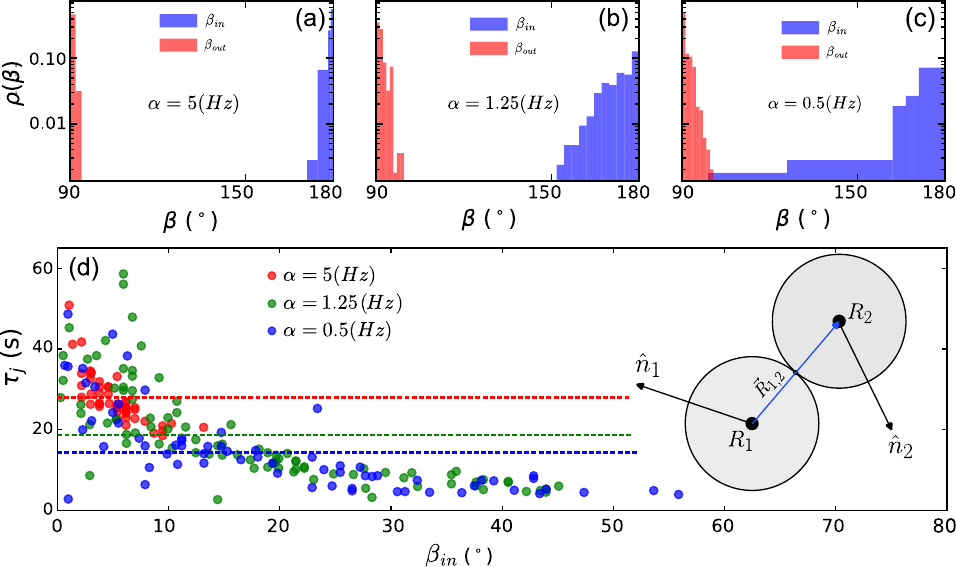}
 \caption{(a–c) Distributions of entry ($\beta_{\mathrm{in}}$) and exit ($\beta_{\mathrm{out}}$) angles, measured upon entering and leaving the collision state, for different reset frequencies $\alpha$.
$\beta_{\mathrm{in}}$ depends strongly on $\alpha$, whereas $\beta_{\mathrm{out}}$ is nearly insensitive.
(d) Collision residence time $\tau_j$ versus $\beta_{\mathrm{in}}$: head-on encounters (small $\beta_{\mathrm{in}}$) are longer-lived and occur more frequently at high $\alpha$. the inset to (d) shows a collision event.
}
\label{fig:incoming_statistics}
\end{figure*}

When two robots operate in the same arena under the resetting protocol, their trajectories may intersect, leading to \emph{collision events} that interrupt forward motion. These interactions introduce an additional timescale associated with resolving mutual blocking, thereby contributing to the total first-passage time and energetic cost. To characterize this effect, we analyzed the geometry of approach and separation for two robots undergoing identical reset dynamics but moving independently between reorientations.

Each robot $i=1,2$ is described by its instantaneous position $\mathbf{r}_i(t)$ and heading direction $\hat{\mathbf{n}}_i(t)$, a unit vector pointing along its propulsion axis. A collision is detected when the inter-robot distance $R_{12}$
falls below a threshold $r_c$ (comparable to the robot diameter), while the robots are approaching one another.

To quantify the geometry of each collision, we define two characteristic angles—an \emph{entry angle} $\beta_{\mathrm{in}}$ and an \emph{exit angle} $\beta_{\mathrm{out}}$—relative to the line-of-centers $\hat{\mathbf{r}} = \mathbf{R}_{1,2}/|\mathbf{R}_{1,2}|$ (see Fig.~\ref{fig:incoming_statistics}).
The definitions are as follows:

\begin{enumerate}
    \item \textbf{Entry angle.} At the instant just before contact ($t_{\mathrm{in}}^-$), the robots are moving toward each other. The entry angle is defined by
    \begin{equation}
        \beta_{\mathrm{in}} = \frac{1}{2}\left[\arccos\!\left(-\hat{\mathbf{n}}_1 \!\cdot\! \hat{\mathbf{r}}\right)
        + \arccos\!\left(\hat{\mathbf{n}}_2 \!\cdot\! \hat{\mathbf{r}}\right)\right].
        \label{eq:beta_in}
    \end{equation}
    With this convention, $\beta_{\mathrm{in}} = 0$ corresponds to a perfectly head-on approach where both headings are directly opposed along the line of centers, and $\beta_{\mathrm{in}} = \pi/2$ represents a grazing approach where both headings are nearly tangent to the contact line.
    \item \textbf{Exit angle.} Upon separation ($t_{\mathrm{out}}^+$), the same construction is applied using the updated heading directions:
    \begin{equation}
        \beta_{\mathrm{out}} = \frac{1}{2}\left[\arccos\!\left(\hat{\mathbf{n}}_1 \!\cdot\! \hat{\mathbf{r}}\right)
        + \arccos\!\left(-\hat{\mathbf{n}}_2 \!\cdot\! \hat{\mathbf{r}}\right)\right].
        \label{eq:beta_out}
    \end{equation}
    Here, $\beta_{\mathrm{out}} = 0$ indicates that both robots depart along the line of centers, while $\beta_{\mathrm{out}} = \pi/2$ corresponds to tangential escape trajectories.
\end{enumerate}

Panels~(a--c) of Fig.~\ref{fig:incoming_statistics} display the distributions of $\beta_{\mathrm{in}}$ and $\beta_{\mathrm{out}}$ for representative reset frequencies~$\alpha$. Increasing~$\alpha$ produces a strong bias toward small~$\beta_{\mathrm{in}}$, indicating that frequent reorientations randomize headings and make head-on encounters more probable. In contrast, the $\beta_{\mathrm{out}}$ distributions remain nearly unchanged with~$\alpha$, suggesting that the exit geometry is determined primarily by the avoidance protocol rather than the pre-collision statistics.

Panel~(d) shows the \emph{collision residence time} $\tau_j$—the duration of the collision—plotted against~$\beta_{\mathrm{in}}$. Head-on collisions (small~$\beta_{\mathrm{in}}$) are both longer-lived and more frequent at high~$\alpha$, whereas grazing encounters (large~$\beta_{\mathrm{in}}$) resolve rapidly through side-passing maneuvers. The inset in Fig.~\ref{fig:incoming_statistics}(d) illustrates a representative head-on event.

These measurements reveal that increasing the reset frequency~$\alpha$ amplifies both the likelihood and duration of head-on collisions. The additional time spent in these prolonged interactions effectively increases the average delay associated with each reset. In the mean first-passage time framework, this manifests as an upward renormalization of the reorientation delay~$\tau_0$, leading to a systematic shift in both the location and magnitude of the minimum in $\langle \tau \rangle(\alpha)$ observed in Fig. 2 of the main paper.